\documentstyle[11pt,aaspp4]{article}
\input epsf.tex
\newcommand{\onu}{\Omega_\nu}
\newcommand{\oc}{\Omega_c}
\newcommand{\Qrms}{Q_{\rm rms-PS}}

\newcommand{\kfs}{k_{\rm fs}}
\begin{document}
\title{Linear Power Spectra in Cold+Hot Dark Matter Models: Analytical 
Approximations and Applications}
\author{Chung-Pei Ma\footnote{e-mail: cpma@tapir.caltech.edu}}
\affil{Theoretical Astrophysics 130-33, California Institute of Technology,
  Pasadena, CA 91125}
\authoremail{cpma@tapir.caltech.edu}
\def\go{\mathrel{\raise.3ex\hbox{$>$}\mkern-14mu
             \lower0.6ex\hbox{$\sim$}}}
\def\lo{\mathrel{\raise.3ex\hbox{$<$}\mkern-14mu
             \lower0.6ex\hbox{$\sim$}}} \def\onu{\Omega_\nu}

\begin{abstract}  
This paper presents simple analytic approximations to the linear power
spectra, linear growth rates, and rms mass fluctuations for both
components in a family of cold+hot dark matter (CDM+HDM) models that
are of current cosmological interest.  The formulas are valid for a
wide range of wavenumber, neutrino fraction, redshift, and Hubble
constant: $k\lo 10\,h$ Mpc$^{-1}$, $0.05\lo \onu\lo 0.3$, $0\le z\lo
15$, and $0.5\lo h \lo 0.8$.  A new, redshift-dependent shape
parameter $\Gamma_\nu=a^{1/2}\onu h^2$ is introduced to simplify the
multi-dimensional parameter space and to characterize the effect of
massive neutrinos on the power spectrum.  The physical origin of
$\Gamma_\nu$ lies in the neutrino free-streaming process, and the
analytic approximations can be simplified to depend only on this
variable and $\onu$.  Linear calculations with these power spectra as
input are performed to compare the predictions of $\onu\lo 0.3$ models
with observational constraints from the reconstructed linear power
spectrum and cluster abundance.  The usual assumption of an exact
scale-invariant primordial power spectrum is relaxed to allow a
spectral index of $0.8\lo n\le 1$.
It is found that a slight tilt of $n=0.9$ (no tensor mode) or $n=0.95$
(with tensor mode) in $\onu\sim 0.1-0.2$ CDM+HDM models gives a power
spectrum similar to that of an open CDM model with a shape parameter
$\Gamma=0.25$, providing good agreement with the power spectrum
reconstructed by Peacock and Dodds (1994) and the observed cluster
abundance.
\end{abstract}
\keywords{cosmology : theory -- dark matter -- elementary particles
-- large-scale structure of universe}

\section{Introduction}
The linear power spectrum of a given cosmological model serves as the
basic input for most linear calculations of large-scale structure and
numerical simulations of gravitational collapse.  The computation of
the power spectrum involves numerical integration of the linearized,
coupled Einstein, Boltzmann and fluid equations that govern the
evolution of the metric perturbations and the density fields of
different particle species.  The calculations become increasingly
complicated as more particle species are included, with those for
massive neutrinos being the most time-consuming due to their
time-dependent energy-momentum relation and nonzero anisotropic stress
(Ma \& Bertschinger 1995 and references therein).

The Boltzmann integration is necessary for obtaining highly accurate
power spectra, but when an accuracy of $\lo 10$\% is all that is
desired, analytic approximations can be very useful and illuminating.
The power spectra for $\Omega=1$ CDM models (with or without a
cosmological constant) have been computed and various fitting
functions have been given (e.g., Bardeen et al. 1986 for zero-baryon
models; Holtzman 1989; Efstathiou, Bond, \& White 1992).  But to date,
no complete spectra have been published for both components in the
CDM+HDM models for all cosmologically interesting ranges of
parameters.  Specific models have been examined, with various
approximations made in the Boltzmann calculation.  Holtzman (1989)
compiled fitting formulas for the present-day {\it baryon} transfer
function for $\onu=0.1$ and 0.3.  Van Dalen \& Schaefer (1992)
tabulated fits to the previously computed transfer functions
(Schaefer, Shafi, \& Stecker 1989) for the density-weighted power
spectrum at $z=10$ for discrete values of $\onu$ between 0 and 0.53.
Klypin et al. (1993) studied the $\onu=0.3, \Omega_b=0.1$ model.
Pogosyan \& Starobinsky (1995) provided a fitting formula for the
linear gravitational potential.  Although the total power spectrum is
useful for many linear calculations, the separate CDM and HDM spectra
and the growth rate are essential, for example, for generation of
initial conditions in numerical simulations.  Moreover, all the
Boltzmann calculations above either assumed zero baryons, which can
lead to discrepancies as large as 25\% for models with 5\% baryons
(see \S 2.2 below), or ignored the anisotropic stress and the higher
moments in the distribution function of the massive neutrinos.

Section 2 of this paper presents analytic approximations to the linear
growth rate of the density field, the CDM and HDM power spectra, the
density-weighted power spectrum, and the rms mass fluctuations in
spatially flat $\onu\lo 0.3$ CDM+HDM models.  The approximations
generally differ by less than 10\% from the numerically integrated
results for the ranges $k\lo 10\,h$ Mpc$^{-1}$, $0\le z\lo 15$, and
$0.5\lo h \lo 0.8$.  The power spectra used in the fit are computed
from the Boltzmann code discussed in Ma \& Bertschinger (1995), which
includes a careful treatment of the massive neutrino phase space and
uses 50 Legendre-modes in the angular expansion of the distribution
function.  The calculations include CDM, baryons, photons, and one
massive and two massless neutrino species.  The range $0\le\onu\lo
0.3$ is chosen since it spans the standard (albeit troubled) pure CDM
model ($\onu=0$) and the previously preferred $\onu=0.3$ model in the
literature.  Late galaxy formation due to excessive neutrino
free-streaming invalidates the $\onu=0.3$ model (Mo \& Miralda-Escude
1994; Kauffmann \& Charlot 1994; Ma \& Bertschinger 1994; Klypin et
al. 1995), leaving $0<\onu< 0.3$ the only open window for the standard
CDM+HDM models.  Numerical simulations of COBE-normalized CDM+HDM
models indicate that for an exact scale-invariant power spectrum
($n=1$), a value of $\onu\approx 0.2$ is needed to simultaneously
satisfy the high-redshift constraint from damped Lyman-$\alpha$
absorbers and the zero-redshift constraint from galaxy pairwise
velocity dispersions (Ma \& Bertschinger 1994; Ma 1995).

In \S 3, linear theory, with the power spectra above as an input,
is used to explore the $\Qrms$, $n$, $h$, and $\onu$ parameter space.
The task of gathering observational constraints and finding a
concordance region in various slices of this multi-dimensional
parameter space has been performed by van Dalen \& Schaefer (1992),
Liddle \& Lyth (1993), Pogosyan \& Starobinsky (1993, 1995), and
Liddle et al. (1995).  The emphasis here is on physical
understanding and explicit illustration of the parameter-dependence of
the growth rates, the power spectra, and cluster abundance.  The
effect of a small tilt ($n\go 0.8$) in the primordial power spectrum
on the cluster abundance is also examined.  Setting the spectral index
$n$ to unity in the standard CDM and CDM+HDM models is, after all,
only a simplification.  Most inflationary models predict {\it nearly}
scale-invariant spectra for the density fluctuations that have an
additional logarithmic dependence on the wavenumber $k$.  The
primordial power spectrum is therefore never exactly scale-invariant,
although a $k$-independent, $n\neq 1$ spectral index approximates the
logarithmic dependence well (Crittenden \& Steinhardt 1992).  The
intention here is to explore small deviations from the canonical
$n=1$, but not to study the models with large tilts that would require
finer tuning of the inflaton potentials.

\section{Analytic Approximations} 

The presence of massive neutrinos in CDM+HDM models complicates the
computation of the power spectrum because the initially relativistic
neutrinos become non-relativistic at $k_B T_\nu \sim m_\nu c^2$.  This
is reflected in the time-dependent energy-momentum relation:
$\epsilon^2=q^2 c^2 + a^2 m_\nu^2 c^4$, where $\epsilon$ is the
comoving energy, $a$ is the expansion factor, and $q$ denotes the
magnitude of the conjugate momentum, which is a constant for
free-streaming neutrinos in the absence of metric perturbations.  As a
result, although the $q$-dependence in the phase space distribution
can be integrated out for massless particles in the Boltzmann
calculation, the distribution of massive neutrinos must be computed on
an additional grid for the momentum space.  This lengthens the
numerical integration time considerably, especially for large
wavenumbers (see, e.g., Ma \& Bertschinger 1995).

The presence of massive neutrinos also considerably complicates the
procedure for fitting the power spectrum because the free streaming of
the neutrinos retards the growth of the density fields, but only below
the free-streaming scale.  The amount of suppression depends on the
redshift, the length scale, the Hubble constant, and the neutrino mass
(or equivalently, $\onu$).  As a result, the linear power spectrum
does not simply grow as $a^2$ on all length scales as in the standard
CDM model, and the analytic approximations will depend on $\onu$, $h$,
and $a$ in addition to the wavenumber $k$.  The simple fitting
procedure to a function $P_c(k)$ of a single variable in the CDM model
therefore becomes a multi-variable problem when massive neutrinos are
present.

Figure~1 shows the numerical results from our Boltzmann
code (available at http://arcturus.mit.edu/cosmics/; see Bertschinger
1995 and Ma \& Bertschinger 1995 for descriptions) for the present-day
($a=1$) power spectra for the pure CDM and four CDM+HDM models.  The
spectral index is taken to be $n=1$, and $H_0=50$ km s$^{-1}$
Mpc$^{-1}$.  One would perhaps first adopt the brute force approach of
fitting the cold and hot spectra $P_c$ and $P_\nu$ directly for the
ranges of $k, \onu$, $a$, and $H_0$ of interest.  However,
Figure~1 shows that unlike $P_c$, $P_\nu$ does not vary
systematically with $\onu$ for $k\go 0.5 h$ Mpc$^{-1}$.  This behavior
makes it difficult to envision a simple functional form to approximate
$P_\nu$.

One solution is to fit the ratios of the power spectra rather than the
spectra directly.  Figures~2 and 3 show the
behavior of three functions $f, g$, and ${\cal H}$ that are defined by
\begin{eqnarray}
     f(k,a,\onu)&=&{1\over 2} {d\log P_c(k,a,\onu)\over d\log a} 
	\,,\nonumber\\
     P_c(k,a,\onu) &=& g(k,a,\onu)\, P_c(k,a,\onu=0) \,,\label{ppp}  \\
     P_\nu(k,a,\onu) &=& {\cal H}(k,a,\onu)\, P_c(k,a,\onu) \,.\nonumber
\end{eqnarray} 
The function $f(k,a,\onu)$ represents the growth rate of the CDM
density field, $g(k,a,\onu)$ represents the ratio of the CDM power
spectrum $P_c$ in CDM+HDM models to that in the pure CDM model, and
${\cal H}(k,a,\onu)$ represents the ratio of the power in the hot and
cold components in a given model.  The merit of using these functions
is that they all vary systematically with $\onu$, $k$, and $a$, as
shown in Figures~2 and 3.  It is therefore
much easier to formulate an elegant ansatz for the analytic
approximations and obtain greater accuracies for the fits.

There are still four parameters $k, \onu$, $a$, and $H_0$ to be
considered.  Further simplification is achieved by identifying
physical processes that introduce special length scales into the power
spectrum.  It is useful to first recall that the power spectrum for a
pure CDM model exhibits a break at $k_{\rm eq}$, the horizon size at
matter-radiation equality.  Since $k_{\rm eq}\propto \Omega_m h^2$ for
a model with a total matter density $\Omega_m$, and the observable
wavenumbers are in units of $h$ Mpc$^{-1}$, the shape parameter
$\Gamma=\Omega_m h$ can be used to characterize this single feature in
the pure CDM power spectrum (Efstathiou et al. 1992).  The
same break should also exist in the power spectra of CDM+HDM models
since the density perturbations in both components cannot grow if
they enter the horizon in the radiation-dominated era.  In this paper,
$\Omega_m=1$, so $\Gamma=h$.  

For the CDM+HDM models, the neutrino free-streaming process introduces
a second length scale in the power spectrum.  The scaling of the
free-streaming distance can be understood in terms of the (comoving)
neutrino Jeans wavenumber $\kfs^2=4\pi G\rho a^2/v^2_{\rm med}$, where
$v_{\rm med}$ is taken to be the median neutrino speed in the
Fermi-Dirac distribution.  (See Bond \& Szalay 1983 for a discussion
of Jeans lengths for collisionless particles.)  For $k<\kfs$, the
density perturbation in the neutrinos is Jeans unstable and grows
unimpeded in the matter-dominated era; for $k>\kfs$, the density
perturbation decays due to neutrino phase mixing.  When the neutrinos
are relativistic, $v_{\rm med} \sim c\,$, and the free-streaming
distance is approximately the particle horizon, which scales as
$\kfs(a) \propto a^{-1}$ (in the radiation-dominated era).  The
neutrinos become non-relativistic at $a\sim 3 k_B T_{0\,\nu}/m_\nu
c^2$ (where $T_{0\,\nu}$ is the neutrino temperature today), after
which $v_{\rm med} \propto k_B T_{0\,\nu}/am_\nu c$.  The relations
$m_\nu\propto\onu h^2$ and $\rho\propto a^{-3} h^2$ then imply
$\kfs(a)\propto a^{1/2}\onu h^3$.  The decrease with time in the
free-streaming distance of non-relativistic neutrinos allows neutrinos
to fall into the CDM potential well and to grow with the CDM on
increasingly smaller scales.  This effect can be seen by comparing $f$
and ${\cal H}$ in Figure~2 and 3, where both
functions remain at unity to a larger $k$ at $a=1$ than $a=0.1$.

Based on the reasoning above, one can simplify the dependence on the
parameters $a,\onu$, and $h$ by introducing a second shape parameter,
\begin{equation}
	\Gamma_\nu=a^{1/2}\onu h^2\,, 
\end{equation}
that characterizes the effect of
decreasing free-streaming distances on the power spectrum.  This
realization was crucial for the simplification of the functional forms
and the improvement of the fits described below.

Below I describe the analytic approximations to the linear growth
rate, the separate CDM and HDM power spectra, the density-weighted
power spectrum, and the rms mass fluctuations.  A summary of the
equations and the fitting coefficients is given in Table 1.


\subsection{Growth Rates}

In the standard CDM model, the CDM density field $\delta_c$ grows as
the expansion factor on all scales; therefore $f\equiv
d\log\delta/d\log a=1$.  The massive neutrinos in CDM+HDM models
introduce an additional length scale, the free-streaming distance,
below which the density fluctuations are washed out and the growth
rate is retarded.  Consequently, the growth rates in CDM+HDM models
are functions of the wavenumber $k$, $\onu$, and time.  This
dependence is illustrated in Figures~2a and 3a.

Before searching for an analytical approximation, it is useful to
ascertain first the asymptotic behavior of the function.  Since HDM
behaves like CDM above the free-streaming distance, $f\rightarrow 1$
as $k\rightarrow 0$.  At large $k$ where the HDM density field is
negligible compared to CDM, the time evolution of the latter is
governed by the linearized fluid equation
$\ddot{\delta_c}+\dot{a}\dot{\delta}_c/a=1.5 H^2 a^2\oc \delta_c$,
where the dots denote differentiation with respect to the conformal
time $\tau$.  Since $Ha=2/\tau$ in the matter-dominated era, the
growing solution is easily shown to give (Bond, Efstathiou, \& Silk
1980)
\begin{equation}
	f_\infty\equiv f(k\rightarrow\infty)
	={1\over 4}\sqrt{1+24\oc}-{1\over 4}
        ={5\over 4}\sqrt{1-{24\over 25}\onu}-{1\over 4}\,.
\label{finfy} 
\end{equation} 
It is interesting to note that
\begin{equation}
	f_\infty \approx \oc^{0.6} 
\end{equation} 
is an excellent approximation to equation~(\ref{finfy}), especially
for the range $\oc\go 0.7$ studied in this paper.  It is sometimes
convenient to express all dependence in terms of $\onu=1-\oc$, and
\begin{equation}
	f_\infty \approx \oc^{0.6} \approx 1-0.68 \onu^{1.05}
\label{finfy2} 
\end{equation} 
is accurate to better than 99.7\% for $\onu\lo 0.3$ (for both
expressions).

It is {\it not} a coincidence that the form above is identical to the
widely-used formula $f=\Omega_m^{0.6}$ for a mildly open model with
matter density $\Omega_m$: equation~(\ref{finfy}) is exact for the
growth rates in open models if $\oc$ is replaced with $\Omega_m$, and
a CDM+HDM model in fact evolves like an open universe with an
effective density $\oc<1$ on scales much below the neutrino
free-streaming distance.  Nevertheless, it should be remembered that
while $f$ is scale-independent for open models, it is scale-dependent
for CDM+HDM models.

An excellent approximation to the growth rate is given by
\begin{equation}
    f(x,\onu)={1+ \oc^{0.6} a_1\,x^{a_2} \over 1+a_1\,x^{a_2}}\,,\qquad
	x\equiv {k\over \Gamma_\nu h} \,,
\label{f}
\end{equation}
where $\Gamma_\nu=a^{1/2}\onu h^2$, $\oc+\onu=1$, and the best-fit
parameters are $a_1=0.1161$ and $a_2=1.363$ for $k$ in units of
Mpc$^{-1}$.  The fractional error of the fit relative to the
numerically computed values is smaller than 0.5\% for $k\lo 20\,h$
Mpc$^{-1}$, $0.05\lo\onu\lo 0.3$, $0.5\lo h\lo 0.8$, and $0\le z \lo
15$.  The functional form is chosen to approach the asymptotic values
discussed above.  As promised, equation~(\ref{f}) depends only on the
variable $x$ that characterizes the neutrino free-streaming scale, and
$\onu$ (or $\oc$) via $f_\infty$.  The solid and dashed curves in
Figures~2a and 3a compare the exact and
fitted growth rates at $a=1$ and 0.1 in different CDM+HDM models with
$h=0.5$.  The top panel of Figure~4 illustrates the
perfect scaling with the Hubble constant through $k/\Gamma_\nu
h\propto k/h^3$.  The shape parameter $\Gamma$ associated with
matter-radiation equality discussed earlier does not affect $f$ in the
matter-dominated era of concern here.

\subsection{CDM Power Spectrum}

Since the strategy here is to use the ratios of the spectra, it is
essential to have an accurate approximation for the power spectrum
$P_c(k,a,\onu=0)$ in the pure CDM model as a starting point.  I will
adopt the functional form of Bardeen et al. (1986)
\begin{equation}
	P_c(q,a,\onu=0) =
	a^2\,A\,k^n \left[{\ln(1+\alpha_1 q)\over \alpha_1 q}\right]^2 
	 {1\over [1+\alpha_2 q+(\alpha_3 q)^2+(\alpha_4 q)^3
	+(\alpha_5\,q)^4]^{1/2}}\,,
	\quad q={ k\over \Gamma h} \,, 
\label{bbks}
\end{equation} 
where $\Gamma=\Omega_m h$ as discussed above.
However, the coefficients $\alpha_1=2.34,\alpha_2=3.89,\alpha_3=16.1,
\alpha_4=5.46$, and $\alpha_5=6.71$ from Bardeen et al. (1986) are accurate 
for the zero-baryon CDM model only.  Increasing the baryonic fraction
$\Omega_b$ decreases the power since the baryonic component cannot
fall into the CDM potential wells until after recombination.  Compared
to the standard CDM model with $\oc=0.95$ and $\Omega_b=0.05$,
equation~(\ref{bbks}) overpredicts the power by as much as 25\% at
$k\go 0.1$ Mpc$^{-1}$ (for fixed long-wavelength normalization $A$).
A high-accuracy fit for the $\Omega_b=0.05$ CDM model can be achieved
by modifying the coefficients to $\alpha_1=2.205,
\alpha_2=4.05, \alpha_3=18.3, \alpha_4=8.725$, and $\alpha_5=8.0$.
The fractional error relative to the direct numerical result is
smaller than 1\% for $k<40\,h$ Mpc$^{-1}$.  Alternatively, replacing
$\Gamma=\Omega_m h$ with $\Gamma=\exp(-2\Omega_b) \Omega_m h$ and using
the Bardeen et al. coefficients in equation~(\ref{bbks}) can
approximate the effect of the baryons (Peacock \& Dodds 1994).  This
enables one to explore a wide range of $\Omega_b$, but the error is
larger for the standard $\Omega_b=0.05$ model: $\lo$ 10\% compared to
$\lo$ 1\% provided by the new coefficients above.

If the temperature fluctuations arise purely from the Sachs-Wolfe
effect (Sachs \& Wolfe 1967), the normalization factor $A$ in
equation~({\ref{bbks}}) is related to $\Qrms$ by
\begin{equation}
	A= k_0^{1-n} {3\over 20\pi} \left({2c\over H_0} \right)^4 
	  {\Qrms^2\over T_{0,\,\gamma}^2}
         = 2.689\times 10^3\,k_0^{1-n}
	\left( {Q_{18}\over T_{2.726}} \right)^2
	\,h^{-4}\,{\rm Mpc}^4 \,,
\end{equation}
where $Q_{18}\equiv \Qrms/18\,\mu K\,$,
$T_{2.726}=T_{0,\,\gamma}/2.726 K\,$, and
$k_0\approx\ell\,H_0/2c\approx 1/600\,h$ Mpc$^{-1}$ for COBE.  There
are other corrections to this formula, but it is good to $\lo$ 1\% for
$\Omega=1$ CDM models with $h=1$ and $\lo$ 4\% for $h$ as low as 0.3
(Bunn, Scott, \& White 1995; White \& Bunn 1995).  The best COBE value
for $\Qrms$ is also model-dependent and is discussed in \S 3.1.

The next step is to relate the CDM power spectra $P_c$ in models with
different $\onu$.  Figures~2b and 3b
illustrate the effect on $P_c$ when CDM is partially replaced by HDM
in a spatially-flat model.  Since the massive neutrinos do not cluster
until the neutrino temperature has dropped below their mass scale, the
CDM component evolves more slowly early on with an effective
$\Omega=1-\onu < 1$ at $k$ above the free-streaming scale .
Consequently, the ratio $g=P_c(k,\onu)/P_c(k,\onu=0)$ becomes smaller
for larger $\onu$.

The large-$k$ behavior of $g$ can be derived analytically from the
definition of the growth rate in equation~({\ref{ppp}).  It is
$g(k\rightarrow\infty)\propto a^{2(f_\infty-1)}$, where $f_\infty$ is
given by equation~(\ref{finfy}).  Using $1-f_\infty
\propto \onu^{1.05}$ as given by equation~(\ref{finfy2}), I find
the following functional form to give a good approximation:
\begin{equation}
  {P_c(k,a,\onu)\over P_c(k,a,\onu=0)} = g(x,\onu)
	= \left( { 1+b_1\,x^{b_4/2}+b_2\,x^{b_4} \over 1+b_3\,x_0^{b_4} }
	\right)^{\onu^{1.05}}\,,
   \quad x={k\over \Gamma_\nu} \,, \quad x_0=x(a=1) \,,
\label{g}
\end{equation} 
where $\Gamma_\nu=a^{1/2}\onu h^2$, and the best-fit parameters are
$b_1=0.01647, b_2=2.803\times 10^{-5}, b_3=10.90$, and $b_4=3.259$
for $k$ in units of Mpc$^{-1}$.  It is important to note that $x$ is
defined to be $k/\Gamma_\nu$ instead of $k/\Gamma_\nu h$ as in
equation~(\ref{f}).  This results from the fact that the two shape
parameters $\Gamma=h$ ($\Omega_m=1$ in this paper) and
$\Gamma_\nu=a^{1/2}\onu h^2$ discussed earlier depend on different
powers of the Hubble constant, and both parameters affect the shape of
$g$, which involves $P_c$ from two different models.  Since the
free-streaming wavenumber $k_{\rm fs}$ is much larger than $k_{\rm
eq}$ at the redshifts of concern here ($z\lo 20$), $\Gamma\propto h$
tends to control the scaling with $h$ in the region where $g$ starts
to deviate from unity in Figures~2 and
3, while $\Gamma_\nu\propto h^2$ controls the scaling
with $h$ at larger $k$.  This mixed dependence on $h$ causes potential
complication for the fitting formula.  For the range $0.5\lo h\lo 0.8$
and $k\lo 10\,h$ Mpc$^{-1}$ of interest here, however, I find that the
scaling $x\propto k/h^2$ works fairly well, as illustrated by the
middle panel of Figure~4.  Therefore, $x$ is modified
to depend on one less power of $h$ than for the function $f$ to
account for the effect of $\Gamma$.  The error in $P_c$ in
equation~(\ref{g}) relative to the numerically computed values is
$\lo$~10\% for $k\lo 10\,h$ Mpc$^{-1}$, $0.05\lo\onu\lo 0.2$, $0.5\lo
h\lo 0.8$, and $0\le z \lo 15$.  For $\onu\approx 0.3$, the error is
still $\lo$~10\% for $h=0.5$ and $k\lo 10\,h$ Mpc$^{-1}$, but it
increases to $\sim 20$\% for $h=0.8$ at high $k$.

\subsection{HDM Power Spectrum}
The HDM to CDM ratio ${\cal H}(k,\onu)$ decreases monotonically with
decreasing $\onu$ as shown by Figures~2c and 3c.  This trend is caused
by the longer neutrino free-streaming lengths for smaller neutrino
masses (hence smaller $\onu$), which makes $\cal H$ deviate from unity
at smaller $k$.  As discussed earlier, the dependence of $P_\nu/P_c$
on $k,a,\onu$ and $H_0$ can be combined in a single variable $x$ that
characterizes the neutrino free streaming distance.  A good analytic
approximation is given by
\begin{equation}
	{P_{\nu}(k,a,\onu)\over P_c(k,a,\onu)} = {\cal H}(x) =
	{e^{-c_1x}\over 1 + c_2 x^{1/2} + c_3 x + c_4 x^{3/2} + c_5x^2} 
		\,,\qquad x\equiv {k\over \Gamma_\nu h} \,,
\label{h}
\end{equation}
where $\Gamma_\nu=a^{1/2}\onu h^2$, and $c_1=0.0015, c_2=-0.1207,
c_3=0.1015, c_4=-0.01618$ and $c_5=0.001711$.
The resulting $P_\nu$ has an error $\lo 15$\% relative to the
numerical results for $k\lo 10\,h$ Mpc$^{-1}$, $0.05\lo\onu\lo 0.3$,
$0.5\lo h\lo 0.8$, and $0\le z \lo 15$.

The bottom panel of Figure~4 illustrates the perfect
scaling of $\cal H$ with $x\propto k/h^3$.  The complication due to
different $h$-scalings in the function $g$ (see \S 2.2) does not arise
here because the physical process associated with shape parameter
$\Gamma=h$ introduces breaks in $P_c$ and $P_\nu$ at the same length
scale (i.e., at $k_{\rm eq}$), and the effect interestingly cancels
out in the ratio $\cal H$.

\subsection{Density-Weighted Power Spectrum}
Although the clustering of the cold and hot components is described by
the separate power spectra $P_c$ and $P_\nu$ discussed above, it
sometimes is useful to have an analytic approximation to the
density-weighted power spectrum $P(k)=\{\onu P_\nu^{1/2} + (1-\onu)
P_c^{1/2}\}^2$ that measures the total gravitational fluctuations
contributed by the separate components.  Here I have assumed that CDM
and baryons have the same power (i.e., $P_c=P_b$), which is a very
good approximation for the range of redshifts and $\Omega_b$ studied
in this paper.
The functional form used for the CDM spectrum $P_c$ in
equation~({\ref{g}}) works well here, and a good approximation is
given by
\begin{equation}
    P(k,a,\onu) = P_c(k,a,\onu=0)
	\left( { 1+d_1\,x^{d_4/2}+d_2\,x^{d_4} \over 1+d_3\,x_0^{d_4} }
	\right)^{\onu^{1.05}}\,,
   \quad x = {k\over \Gamma_\nu} \,, \quad x_0=x(a=1)\,,
\label{pave}
\end{equation} 
where $P_c(k,a,\onu=0)$ for the pure CDM model is given by
equation~(\ref{bbks}), and $d_1=0.004321, d_2=2.217\times 10^{-6},
d_3=11.63$, and $d_4=3.317$.  The error in the resulting $P$ relative
to the numerically computed values is $\lo$ 10\% for the ranges $k\lo
20\,h$ Mpc$^{-1}$, $0.05\lo\onu\lo 0.3$, $0.5\lo h\lo 0.8$, and $0\le
z \lo 15$.

Pogosyan \& Starobinsky (1995) introduced a more complicated
8-parameter fitting formula for the gravitational potential that gives
results similar to ours. The analytic approximation provided by
Efstathiou et al. (1992), however, does not fit the
density-weighted CDM+HDM $P(k)$ well.  Their formula
$P(k) = Bk /\{1 + [ak + (bk)^{3/2} + (ck)^2]^\nu\}^{2/\nu}\,$,
with $a=6.4/(\Gamma h)$ Mpc, $b=3.0/(\Gamma h)$ Mpc, $c=1.7/(\Gamma
h)$ Mpc, $\nu=1.13$, and $\Gamma=0.2(\onu/0.3)^{-0.5}$ does not take
into account the redshift dependence of the shape and does
not have the correct dependence on $\onu$ at large $k$.  It produces
too little power at $0.01\lo k \lo 1$ Mpc$^{-1}$ by as much as a
factor of $\sim$ 2.
On the other hand, this formula with $\Gamma=\Omega_m h=0.5$ works
fairly well for the $\Omega_b=0.05, h=0.5$ flat CDM model.  The
maximal discrepancy is $\sim$ 18\%, compared to 25\% in the original
Bardeen et al. fits.

\subsection{Mass Fluctuations $\sigma$}
The linear rms mass fluctuations in spheres of radius $R$ are related
to the power spectrum $P$ by
\begin{equation}
 \sigma^2(R,\onu,a)=\int_0^\infty {dk\over k}\, 4\pi k^3 
	P(k,\onu,a) W^2(kR)\,,
\label{sigma}
\end{equation}
where $W(x)=3(\sin x - x \cos x)/x^3$ is the tophat window function,
and the mass enclosed in the sphere is $M=4\pi\rho_0 R^3/3$, with
$\rho_0$ denoting the background mass density of the universe.
Although different filter windows have been used in the literature,
and there is no {\it a priori} reason to prefer a particular window
function, Lacey \& Cole (1994) find that the tophat gives the best fit
to $N$-body simulations with initial power-law $P(k)$.  More
importantly, the best-fit $\delta_c$ for a tophat is insensitive to the
spectral index $n$.  In contrast, they find the best fit $\delta_c$ to
depend more strongly on $n$ when the Gaussian window is used.  Ma
\& Bertschinger (1994) also find the shapes of the
Press-Schechter mass functions computed with the tophat filter to give
a better match in CDM+HDM simulations.  Since a family of models with
different spectral shapes at large $k$ is being studied here, the
tophat window will be the most robust filter.

Figure~5 shows the present-day $\sigma(R,\onu)$ for $n=1$
and $h=0.5$ models with different $\onu$.  At large $R$, all curves
converge because the same 4-year COBE normalization ($\Qrms=18\,\mu
K$) is used, and $\sigma \propto R^{(-3-n)/2}$ since $P(k)\propto k^n$
at small $k$.  The small-$R$ behavior simply reflects the effect of
neutrino free streaming on the power spectrum.  I will again fit to
$\sigma(\onu=0)$ for the pure CDM model and then the ratios
$\sigma(\onu)/\sigma(\onu=0)$.  An excellent analytic approximation to
$\sigma(\onu=0)$ is given by
\begin{equation}
    \sigma(R,\onu=0)=Q_{18}\,\,h^2 \left( {R\over R_0} \right)^{(1-n)/2}\,
   {1 \over a_1+ a_2\,x^{a_4} + a_3\,x^2 }\,, \qquad x\equiv {R h^2} \,,
\label{sigfitc}
\end{equation}
where $Q_{18}=\Qrms/18\,\mu K, a_1=0.01359, a_2=0.05541, a_3=0.001702,
a_4=0.8032$, $R_0=1000$ Mpc, and $R$ is in units of Mpc.  The exponent
$(1-n)/2$ is chosen so that $\sigma \propto R^{(-3-n)/2}$ at large
$R$.  The fractional error is $\lo$ 5\% for $R=0.07$ -- 350 $h^{-1}$
Mpc, or $M=4\times 10^8$ -- $5\times 10^{19} h^{-1} M_\odot$.  For
CDM+HDM models, one can use
\begin{equation}
    \sigma(R,\onu)=\sigma(R,\onu=0)\, {1+ \oc^{a_3} a_1\,x^{a_2} 
	\over 1+a_1\,x^{a_2}}\,,\qquad x\equiv {R \Gamma_\nu} \,,
\label{sigfit}
\end{equation}
where $\Gamma_\nu=\onu h^2$, $\oc=1-\onu$, and the best-fit parameters
are $a_1=0.7396, a_2=-0.8927$, and $a_3=5.106$.  The fractional error
is $\lo$ 10\%.  The result of the fitting is shown in
Figure~5 (as dashed curves).

As emphasized earlier, the shape of $P(k,a,\onu)$ has a non-trivial
dependence on the redshift due to neutrino free streaming.  However,
the mass fluctuations obey $\sigma(R,\onu,a)=a\,\sigma(R,\onu,a=1)$ to
a good approximation, with $<$10\% error for $M\go 10^{10} M_\odot$
(see Figure~5).  I therefore will not give a
time-dependent approximation.  If $\sigma$ is required to higher precision,
or at a lower mass, one can carry out the simple integral in
equation~(\ref{sigma}) using the analytic approximation for
$P(k,a,\onu)$ in equation~(\ref{pave}).

\section{Observational Constraints}
\subsection{Reconstructed $P(k)$}
The theoretical linear power spectra computed above can be put to the
test against the observed power spectra of galaxies and clusters, if
the effects of non-linear evolution, redshift-space distortions, and
bias are corrected for.  This is not a straightforward task,
especially on highly nonlinear scales.  By limiting the study to the
quasi-linear regime ($k\lo 0.5\,h$ Mpc$^{-1}$), Peacock \& Dodds
(1994) attempted to reconstruct the linear mass power spectrum from
eight independent sets of data from optical, radio, and IRAS galaxies
and optical and radio Abell clusters.  They were able to obtain good
internal agreement for the estimated power spectrum given specific
conditions: a significant redshift-space distortion; a
scale-independent bias for a given class of objects, with relative
bias factors for Abell clusters, radio galaxies, optical galaxies, and
IRAS galaxies required to be 4.5 : 1.9 : 1.3 : 1 to within 6\% rms.
Among the pure CDM models parameterized by the shape parameter
$\Gamma=\Omega_m h$ (see eq.~[\ref{bbks}]), the best fit was found to
be given by $\Gamma\approx 0.25$.

Figure~6 compares the final estimate for the linear
power spectrum from Peacock \& Dodds (1994) with those for the
standard CDM and three CDM+HDM models with $h=0.5$ computed in the
previous section.  (The density-averaged $P=\{\onu P_\nu^{1/2} +
(1-\onu) P_c^{1/2}\}^2$ is used for CDM+HDM models.)  Three primordial
spectral indices $n=1$, 0.95, and 0.9 are shown.  The bottom two
panels illustrate the effect of tensor contributions in the tilted
models.  Whether a model produces a non-zero tensor fluctuation
depends on the shape of the inflaton potential at the time of the
first horizon crossing (60 or so e-folds before the end of inflation).
Inflationary models, such as the extended or chaotic models, generally
give a tensor-to-scalar ratio of $T/S\approx 7(1-n)$ (Davis et al.
1992), although the so-called natural inflationary model predicts a
negligible tensor contribution (Adams et al. 1993).  Since COBE
measures the combined scalar and tensor anisotropies, the effect of a
non-zero tensor mode, when the tilt is small, is to lower the
normalization for the scalar power spectrum by $S/(S+T)$ for a given
$\Qrms$.  

An additional effect on the normalization that must be considered is
the dependence of $\Qrms$ on parameters such as $n$, $\Omega_b$ and
$h$ assumed in a given model.  The dependence arises because there
exist small corrections to the Sachs-Wolfe effect (Sachs \& Wolfe
1967) that is the main contributor to the angular power spectrum for
the temperature fluctuations in the cosmic microwave background on the
COBE scale.  A detailed analysis of such dependence has been carried
out by Bunn et al. (1995), who obtain $\Qrms(n)=21.1
\exp[0.69(1-n)]\,\mu K$ from the 2-year COBE data (for a pure Sachs-Wolfe
spectrum).  A maximum likelihood analysis of the same data by Gorski
et al. (1994) finds $\Qrms(n)=(39+2n) \exp(-0.73n)$ with quadrupole,
and $\Qrms(n)=(40+2n)\exp(-0.74n)$ without quadrupole.  Results from
the 4-year COBE data have been announced recently (Gorski et al. 1996;
Bennett et al. 1996).  I therefore will use the latest value
$\Qrms=18\,\mu K$ for $n=1$ (Gorski et al. 1996), and $\Qrms=
19.2\,\mu K$ and $20.5\,\mu K$ for $n=0.9$ and 0.8 (Gorski, private
communication), respectively.  (Note that using $\Qrms(1)=18$ instead
of 21.1 in the Bunn et al. formula gives very similar results:
$\Qrms=19.3\,\mu K$ for $n=0.9$ and $20.7\,\mu K$ for $n=0.8$.)  The
dependence on $n$ for $n=0.8-1$ is weak, but it is taken into account
in Figure~6.  Varying $\Omega_b$ from 0.01 to 0.1 in
the CDM model changes $\Qrms$ by only about 1\% (Bunn et al. 1995), so
it will not be considered here.

As Figure~6 illustrates, the 4-year COBE result
is consistent with the amplitude of reconstructed power
spectrum at small $k$, but all $n=1$ models with $0\le\onu\le 0.3$
normalized to this value have too much power at $0.05\lo k\lo 0.2
\,h\,$Mpc$^{-1}$.  A slight tilt of $n=0.9-0.95$ brings
the CDM+HDM models into excellent agreement with the data.  For a
given COBE normalization, the power spectrum of a tilted model with a
negligible $T/S$ (bottom panel of Fig.~6) has a higher
amplitude than that of a model with $T/S\approx 7(1-n)$ (middle
panel); therefore a slightly smaller $n$ is generally needed to match
the reconstructed power spectrum.  Figure~6 shows that
the four highest $k$ points in Peacock
\& Dodds (1994) favor $\onu\approx 0.1-0.2$, although Liddle et al.
(1995) recently questioned the validity of these high-$k$ points.  It
is also interesting to note that the spectra for the $\onu=0.1$ model
with $n=0.95$ (with tensor) and $n=0.9$ (without tensor) are very similar
to the open CDM spectrum with $\Gamma=0.25$.

Figure~7 shows the effect of increasing $h$ to 0.65.
The discrepancy worsens for $n=1$, but a larger tilt of $n\approx 0.9$
(with tensor) or $n\approx 0.8$ (without tensor) can compensate for
the increased power and bring the models with $\onu\sim 0.1-0.3$ into
good agreement with the data points.

\subsection{$\sigma_8$ and Cluster Abundance}
The rms linear mass fluctuation on scales of 8$h^{-1}$ Mpc,
$\sigma_8$, has played a very special role in the literature.  Because
the observed rms galaxy count on this scale is about unity (Davis \&
Peebles 1983), $\sigma_8$ was used to define the ``bias factor'',
$b=\sigma_8^{-1}$, between mass and galaxies.  It was also a common
practice to use $\sigma_8$ for the normalization of the linear matter
power spectrum, a value undetermined by most theories.  After COBE,
the rms quadrupole inferred from the cosmic microwave background
anisotropy is often used to fix the normalization of the (matter)
power spectrum.  The corresponding $\sigma_8$ in a given cosmological
model then depends on the shape of the power spectrum.  Analytic
approximations to $\sigma(R)$ have already been given by
equations~(\ref{sigfitc}) and (\ref{sigfit}).  Using $R=8 h^{-1}$ Mpc
gives the following approximation for $\sigma_8$:
\begin{eqnarray}
  &&\sigma_8(\onu=0) = Q_{18}\,{h^2\, 0.008^{(1-n)/2}
       \over 0.0136 + 0.294\,h^{0.803} + 0.109 h^2 }\,, \nonumber\\
  &&\sigma_8(\onu) = \sigma_8(\onu=0)\, {1+ \oc^{5.11}\,0.116\,
	(\onu\,h)^{-0.893} \over 1+ 0.116\,(\onu\,h)^{-0.893} }\,,
\end{eqnarray}
where $Q_{18}=\Qrms/18\,\mu K$.  

The masses and abundances of rich clusters of galaxies provide an
independent constraint on $\sigma_8$.  Evrard (1989) obtained
$\sigma_8\sim 0.5-0.7$ in the standard CDM model from high velocity
dispersion clusters.  Henry \& Arnaud (1991) obtained
$\sigma_8=0.59\pm 0.02$ for scale-free spatially-flat models from
their X-ray cluster temperature function.  White, Efstathiou and Frenk
(1993) found $\sigma_8\sim 0.52-0.62$ for the standard CDM model using
the Henry \& Arnaud cluster data and the Press-Schechter (1974)
approximation (assuming a linear overdensity of $\delta_c=1.68$).
Liddle et al. (1996) used the same data and also took into account the
redshift dependence in the relation between cluster virial masses
and temperatures.  They considered both open and flat CDM models and
found $\sigma_8=0.59^{+.21}_{-.16}$ after allowing $1.6\le \delta_c\le
1.8$.

Here I use the Press-Schechter approximation (1974) to investigate the
constraint imposed on the neutrino fraction $\onu$ in CDM+HDM models
by the cluster data.  The comoving number density of objects of mass
above $M$ at redshift $z$ is given by
\begin{equation}
        N(M,z) = \int_{\ln M}^\infty d\ln M^\prime \sqrt{2\over\pi}
        {\rho_0(z)\over M^\prime}{\delta_c \over \sigma(M^\prime,z)}
        \left| {d\ln\sigma \over d\ln M^\prime} \right|
        \exp \left[-{\delta_c^2 \over 2\sigma^2(M^\prime,z)}\right] \,,
\label{ps}
\end{equation}
where $\rho_0(z)$ denotes the background mass density of the universe,
and $\delta_c$ is a free parameter characterizing the linear
overdensity at the onset of gravitational collapse.  The dependence on
cosmological models is via $\sigma(M)$, which is given by
equations~(\ref{sigfitc}) and (\ref{sigfit}) with $M=4\pi
R^3\,\rho_0/3$.

White et al. (1993a) approximate $\sigma(M)$ in the CDM model as a
single power-law for the cluster mass range.  As Figure~5
illustrates, this is not a good approximation for CDM+HDM models
because $\sigma(M)$ turns over more sharply around $10^{14}$--$10^{15}
M_\odot$ due to the smaller power in $P(k)$.  Liddle et al. (1995, 1996)
adopted an improved approximation as a function of $\Gamma=\Omega_m h$
for open and flat CDM models, but they did not give a formula for
$\sigma(\onu)$ for CDM+HDM models.  To compute the number density
$N(M)$, one should use either $\sigma$ computed from $P(k)$ in
equation~(\ref{pave}) or the analytic $\sigma$ given by
equations~(\ref{sigfitc}) and (\ref{sigfit}) above.

Figure~8 shows the present-day number density of
cluster-scale objects as a function of mass for various models.
Dependence of the cluster number density on the overdensity
$\delta_c$, normalization $\Qrms$, and spectral index $n$ is also
illustrated.  The sensitivity to the parameters arises because cluster
masses reside in the Gaussian tails of the Press-Schechter curves.
Our experience with $N$-body simulations is that the best-fit
$\delta_c$ can range from 1.68 to 1.8 (Ma \& Bertschinger 1994), so
both values are shown for comparison.

Figure~9 illustrates two effects that would help to
lower the predicted $N(>M)$: decreasing $\Qrms$ or increasing
$\Omega_b$.  The value $\Qrms=16.4\,\mu K$ is the lower 1-$\sigma$
normalization given in Bennett et al. (1996).  The high value of
$\Omega_b=0.1$ (for $h=0.5$) is preferred if the recent measurement of
the deuterium to hydrogen ratio in a quasar absorption line system (Tytler,
Fan, \& Burles 1996) is accurate and reflects the primordial value.
Another possibility not studied here is dividing the contribution of
massive neutrinos to $\onu$ among two or more species (Primack et al.
1995).  For a fixed $\Omega_c$, the longer free streaming
distance of the lighter neutrinos helps to lower the power spectrum
slightly and therefore gives a lower $N(>M)$.

The two data points in Figures~8 and
9 indicate the number densities of clusters with
X-ray temperatures exceeding $k_B T=3.7$ and 7 keV from Henry \&
Arnaud (1991).  The mass range shown for the $k_B T > 3.7$ keV
clusters is taken from White et al.  (1993a), where the upper limit
$5.5\times 10^{14} h^{-1} M_\odot$ was estimated from cluster velocity
dispersions assuming an isotropic distribution, while the lower limit
$4.2\times 10^{14} h^{-1} M_\odot$ was converted from the X-ray
temperature assuming an isothermal gas and a density profile of
$\rho\propto r^{-2}$.  Both values correspond to masses within one
Abell radius 1.5 $h^{-1}$ Mpc and are extrapolated from the observed
values at $\sim 0.5 h^{-1}$ Mpc.  The mass range plotted for the $k_B
T > 7$ keV clusters is from the more recent analysis of Liddle et al.
(1996), who used the hydrodynamical simulations by White et al.
(1993b) to calibrate the cluster masses within one Abell radius for a
given X-ray temperature.  Using the density profile $\rho \propto
r^{-2.4}$ (White et al. 1993b), Liddle et al. then converted the mass
above to the virial mass $(1.2^{+0.7}_{-0.5}) \times 10^{15} h^{-1}
M_\odot$ for a cluster with a mean X-ray temperature of 7 keV in a
critical-density universe.

In comparison with these published values, Figure~8
indicates that all $\Qrms=18\,\mu K$, $n=1$ models with $\onu\lo 0.3$
and $\Omega_b=0.05$ overpredict the cluster number density for
$\delta_c=1.68$--1.8.
Models with $n=0.9-0.95$ agree better, with the preferred range of
$\onu$ dependent on the value of $\delta_c$ and whether or not there
is a tensor contribution.  Figure~8 shows the same
tilted models as in Figure~6.  For $n=0.9$ and no
tensor mode (normalized to $\Qrms=19.2\,\mu K$), the preferred
range is $0.2\lo\onu\lo 0.4$, or $0.65\lo \sigma_8\lo 0.8$, for
$\delta_c=1.68$ -- 1.8.  For $n=0.95$ and $T/S=7(1-n)$ (with
$\Qrms=18.5\,\mu K$), the allowed range is $0.1\lo\onu\lo 0.3$, or
$0.6\lo\sigma_8\lo 0.7$.  This preference for $n\approx 0.9-0.95$ is
consistent with the conclusion drawn from the reconstructed power
spectrum in the previous section.  A lower $\Qrms$ or higher $\Omega_b$
in $n=1$ models also helps to decrease the cluster number density,
but the effects are smaller (see Fig.~9).

It should be kept in mind that the mass ranges adopted by White et al.
(1993a) and Liddle et al. (1995, 1996) are obtained under specific
assumptions.  Cluster mass determination is by no means a settled
issue.  Temperature gradients, substructure, and ellipticity can all
affect the X-ray mass estimates (see, e.g., Tsai, Katz, \&
Bertschinger 1994).  Cluster mass function has also been estimated
from velocity dispersions of cluster galaxies (e.g., Biviano et al.
1993), but inclusion of substructure in the analysis lowers the
normalization (Bird 1995).  We can invert the problem and ask what
cluster mass is needed for the currently favored $n=1$ CDM+HDM models
to produce the observed cluster abundance.  Within the narrow window
$\onu\sim 0.1$ -- 0.2 allowed by both the low- and high-redshift
constraints (Ma \& Bertschinger 1994; Ma 1995),
Figure~8 indicates that the 3.7 and 7 keV clusters
would have to have a median mass of $\sim 1-1.5\times 10^{15}$ and
$\sim 2-4\times 10^{15} h^{-1} M_\odot$, respectively, for $\onu=0.1$
-- 0.2 models ($n=1$) to produce the observed cluster abundance.  This
seems very unlikely, given that the hottest and most luminous X-ray
cluster Abell 2163 ($k_B T=12-15$ keV) is recently estimated to have a
mass of only $M_A=1.07\pm 0.13 \times 10^{15} h^{-1} M_\odot$ out to
the Abell radius (Markevitch et al.  1996).  For a power-law density
profile $\rho(r)\propto r^{-\alpha}$ at the outer part of the cluster,
this translates to a virial mass of $M_V\approx 1.5^{(3/\alpha-1)}
M_A$, that is, only about $1.2-1.35
\times 10^{15} h^{-1} M_\odot$ for $\alpha$ ranging from 2.4 from
simulations (White et al. 1993b) to the isothermal $\alpha=2$.

\section{Conclusions}
This paper presents numerical results and analytical approximations to
the linear power spectrum and the density field growth rate for a
class of CDM+HDM models with neutrino fraction $\onu\lo 0.3$ (neutrino
mass $m_\nu\lo$ 7 eV) that are currently of much cosmological
interest.  The models with $\onu=0$ (i.e. the standard CDM) and
$\onu\go 0.3$ have difficulty explaining various observations, leaving
the range of the neutrino masses studied in the paper as the only hope
for the standard CDM+HDM models.  The analytical functions given by
equations (\ref{f})-(\ref{pave}) and (\ref{sigfitc}) and
(\ref{sigfit}) in this paper give accurate approximations to the
linear growth rate, the separate cold and hot power spectra, the
density-weighted power spectrum, and the rms mass fluctuations for
$k\lo 20\,h$ Mpc$^{-1}$, $z\lo 15$, and $0.5\lo h\lo 0.8$ in CDM+HDM
models with $\onu\lo 0.3$.  They should be useful as the input for
most linear calculations and for the initial conditions of numerical
simulations of structure formation in these models.

It is shown that the difference in the growth rate and the power
spectrum between the CDM+HDM and the standard CDM models can be
entirely characterized by the free-streaming distance of the
neutrinos.  A shape parameter $\Gamma_\nu=a^{1/2}\onu\,h^2$ is
introduced to explain the behavior of the power spectra in
Figure~1; it also helps to greatly simplify the functional
forms of equations~(\ref{f}), (\ref{g}), and (\ref{h}), and to improve
the accuracy of the fits.

Linear calculations with these power spectra as input are performed to
study the dependence of cluster abundance on the spectral index $n$,
normalization $\Qrms$, tensor-to-scalar ratio $T/S$, baryon fraction
$\Omega_b$, and neutrino masses in $\onu\lo 0.3$ models.  The
theoretical predictions are compared to the reconstructed $P(k)$ by
Peacock \& Dodds (1994) and the X-ray cluster abundance and masses
from Henry \& Arnaud (1991).  These tests are chosen because the
linear theory (with proper calibration for the clusters) should be
adequate on such large scales.  When normalized to the 4-year COBE
$\Qrms=18\,\mu K$, all $n=1$, $\Omega_b=0.05$, $\onu\lo 0.3$ models
predict too much power by up to a factor of $\sim 2$ when compared to
Peacock \& Dodds (1994) and Herny \& Arnaud (1991).  A lower $\Qrms$
or higher $\Omega_b$ both helps to reduce the discrepancy.  Also
considered is a slight tilt of $n=0.9-0.95$ in the power spectrum,
which brings the theoretical predictions into good agreement with
data.

Due to the heterogeneous nature of the Peacock \& Dodds sample and the
uncertainties in cluster abundance and mass, it is difficult to attach
a meaningful statistical significance to the validity of the various
models studied in this paper.  It should also be kept in mind that the
conclusion that the $n=0.9-0.95$ models provide a closer match to the
data depends on several implicit assumptions.  These include the
scale-independent bias used in Peacock \& Dodds (cf. Cen
\& Ostriker 1992) and the isothermal and spherical model for clusters
used in White et al. (1993a) and Liddle et al. (1996) to convert
X-ray temperature into virialized mass.  These are obviously
simplifications, but modeling the deviations from such assumptions
would require the introduction of parameters that can not be well
determined by current data.  Nonetheless, the theoretical calculations
carried out in this paper should provide useful predictions that can
be tested by future data.

\acknowledgments
The author is grateful to Ed Bertschinger and Peter Goldreich for
valuable discussions, Una Hwang, Mark Metzger, HouJun Mo, Paul
Steinhardt, and Ned Wright for helpful comments, and K. Gorski and
Richard Mushotzky for useful information.  Parts of the Boltzmann
calculations were performed at the National Center for Supercomputing
Applications.  Support from a PMA Division Fellowship at Caltech is
acknowledged.

\clearpage 
\begin{deluxetable}{cclllll}  
\tablecaption{Summary of Analytic Approximations}
\tablewidth{0pt}
\tablehead{\colhead{Symbol} &  \colhead{Equation}   & \colhead{Coefficients 1}
	 & \colhead{2} & \colhead{3} & \colhead{4} & \colhead{5} }
\startdata
  $f$ &	(\ref{f}) & 0.1161 & 1.363 &&& \nl
  $P_c(\onu=0)$  & (\ref{bbks})\tablenotemark{a}   &  
	2.34&3.89& 16.1& 5.46& 6.71  \nl
    & (\ref{bbks})\tablenotemark{b}   &  
	 2.205& 4.05 & 18.3 & 8.725 & 8.0 \nl
$P_c(\onu)$  & (\ref{g})  & 0.01647& 2.803e-5& 10.9& 3.259& \nl
$P_\nu(\onu)$   &  (\ref{h})  & 0.0015& $-0.1207$ & 0.1015& $-0.01618$
	& 0.001711 \nl
$P(\onu)$ & (\ref{pave})\tablenotemark{c}  &
	0.004321& 2.217e-6& 11.63 &3.317 & \nl
$\sigma(\onu=0)$ & (\ref{sigfitc}) & 0.01359 &0.05541& 0.001702 &0.8032& \nl
$\sigma(\onu)$ & (\ref{sigfit}) & 0.7396 & $-0.8927$ & 5.106 & &\nl
\enddata

\tablenotetext{a}{For zero baryons, or modify variable $\Gamma$ in 
	eqn.~(\ref{bbks}) for general baryon fraction (see text)}
\tablenotetext{b}{High accuracy fit (with $\lo$ 1\% error)
	for the standard CDM model with 5\% baryons}
\tablenotetext{c}{Density-weighted spectrum $P=\{\onu P_\nu^{1/2} + 
	(1-\onu) P_c^{1/2}\}^2$ }
\end{deluxetable}

\clearpage
\begin{figure}
\epsfxsize=6.5truein 
\epsfbox{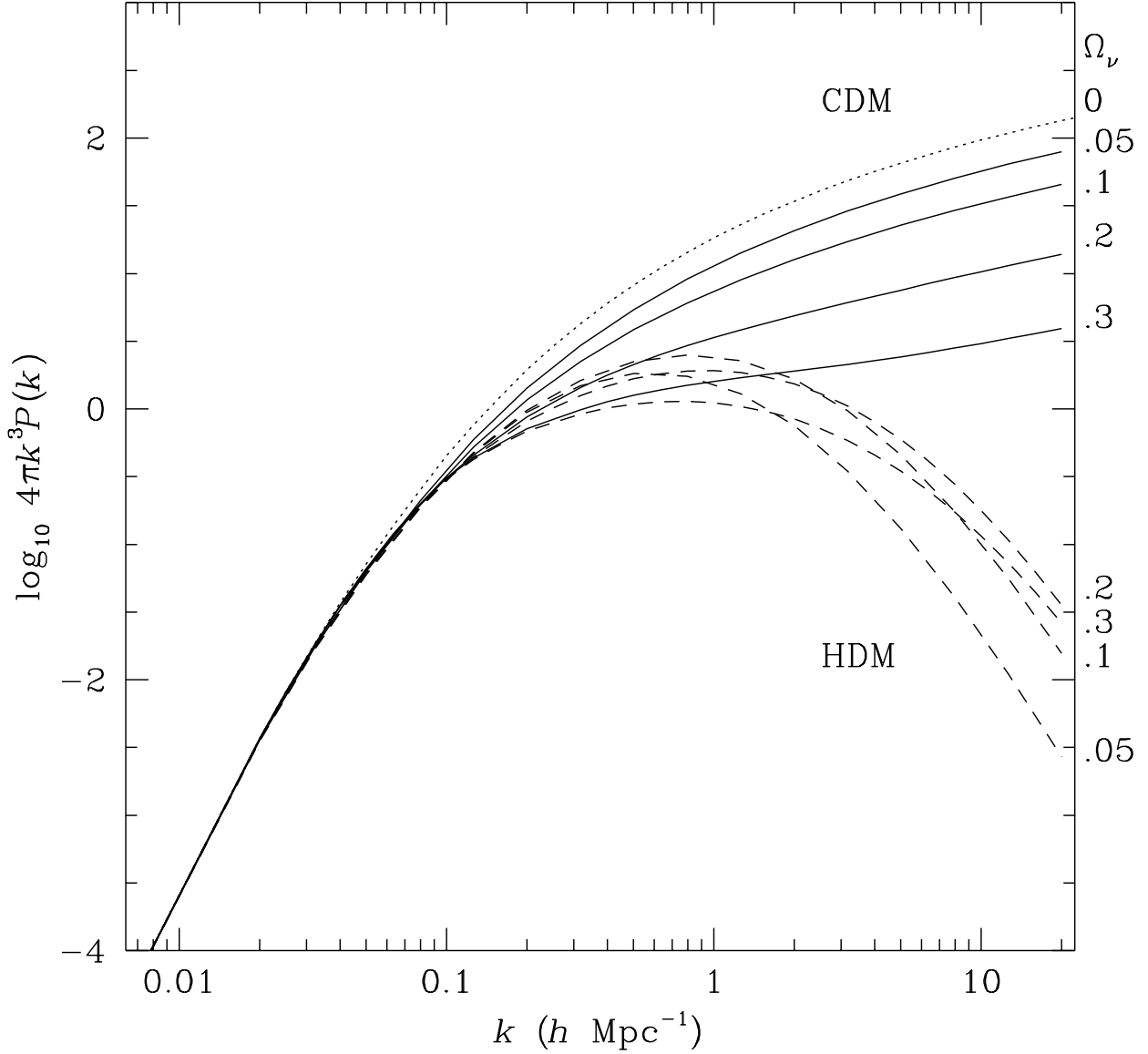}
\caption{Present-day linear power spectra for the standard CDM model
(dotted) and four CDM+HDM models with $n=1$, $\Omega_b$, and $H_0=50$
km s$^{-1}$ Mpc$^{-1}$, computed from integrations of the Boltzmann
equations.  The four CDM+HDM models have $\onu=0.05, 0.1, 0.2$, and
0.3, and the power in the cold (solid) and hot (dashed) components are
shown separately.  All are normalized to the 4-year COBE result
$\Qrms=18\,\mu K$ (Gorski et al. 1996).}
\label{fig:pow}
\end{figure}
\begin{figure}
\epsfxsize=6.5truein 
\epsfbox{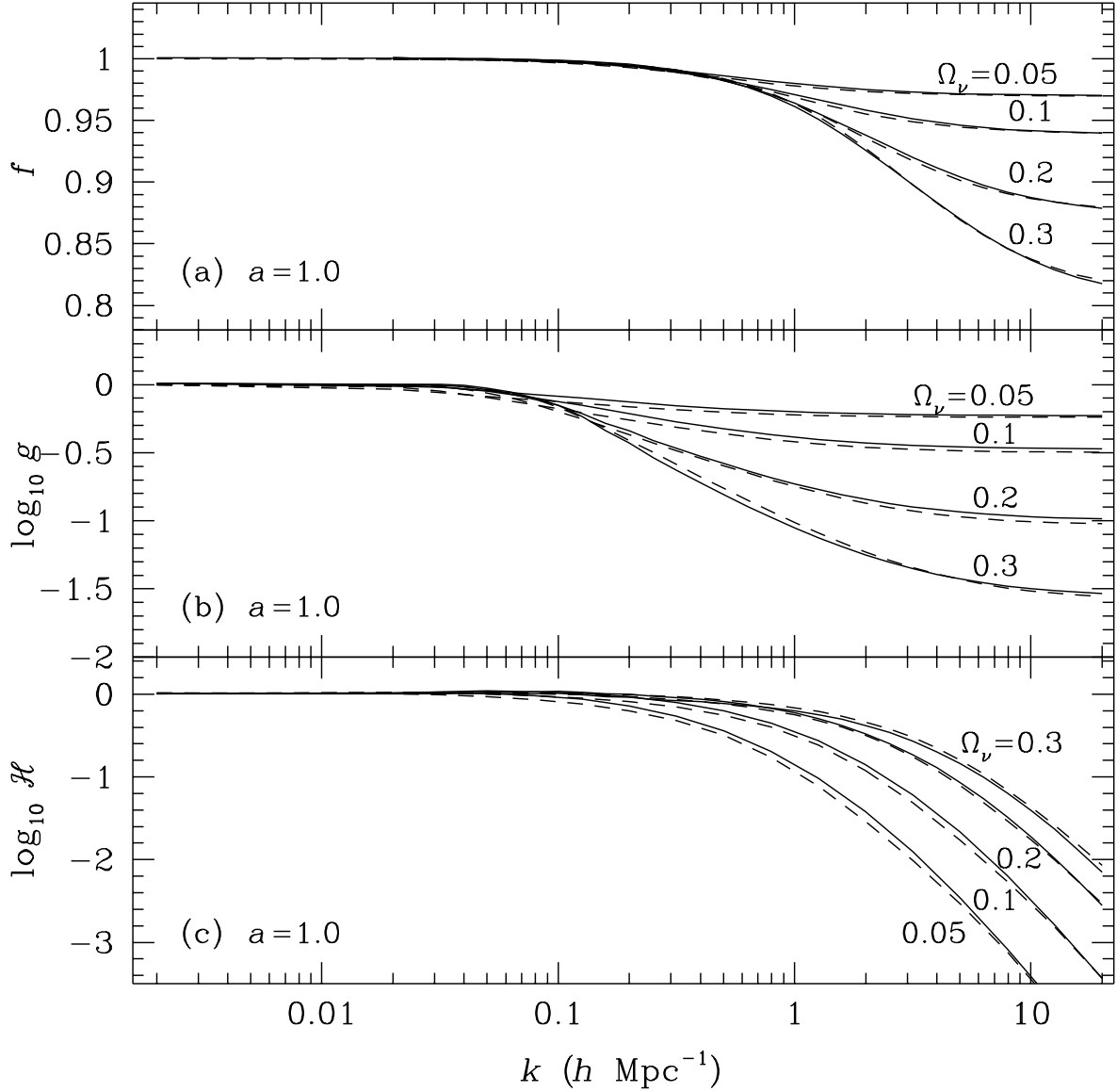}
\caption{Direct integration of the Boltzmann equations (solid)
vs. analytic fitting results (dashed) for the present-day CDM and HDM
power spectra and the growth rate in four CDM+HDM models ($n=1,
H_0=50$) with $\onu=0.05, 0.1, 0.2$, and 0.3.  (a) Growth rate
$f(k,a,\onu)$ of the density field.  The fitting formula is given by
eq.~(\protect{\ref{f}}).  (b) Ratio of the CDM power spectrum in
CDM+HDM models to that in the pure CDM model,
$g=P_c(\onu)/P_c(\onu=0)$.  The fitting formula is given by
eq.~(\protect{\ref{g}}). (c) Ratio of the HDM to CDM power
spectrum, ${\cal H}=P_\nu(\onu)/P_c(\onu)$.  The fitting formula is
given by eq.~(\protect{\ref{h}}). }
\label{fig:fgh1}
\end{figure}
\begin{figure}
\epsfxsize=6.5truein 
\epsfbox{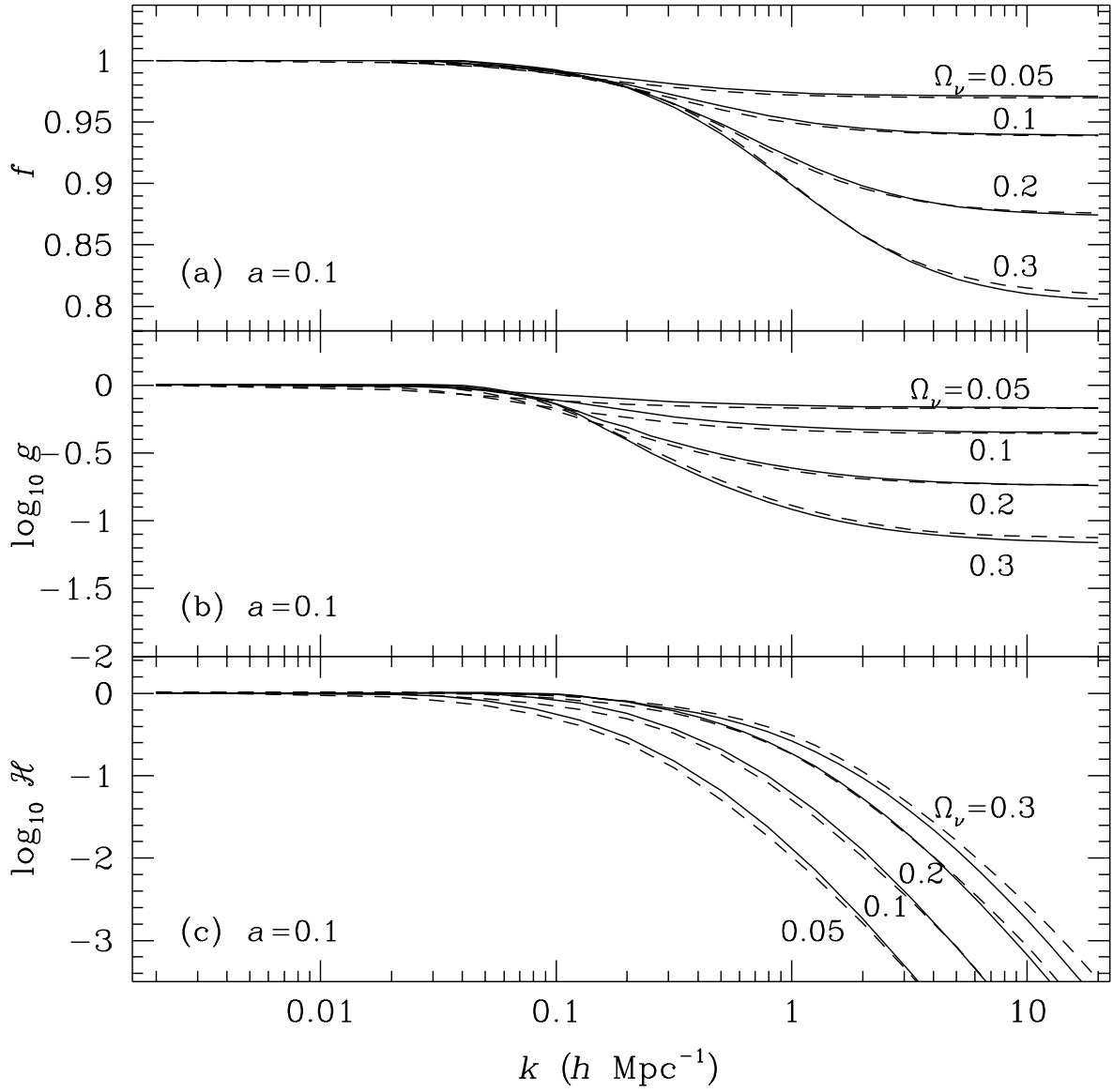}
\caption{Same as Figure~\protect{\ref{fig:fgh1}} but for $a=0.1$}
\label{fig:fgh2}
\end{figure}
\begin{figure}
\epsfxsize=6.5truein 
\epsfbox{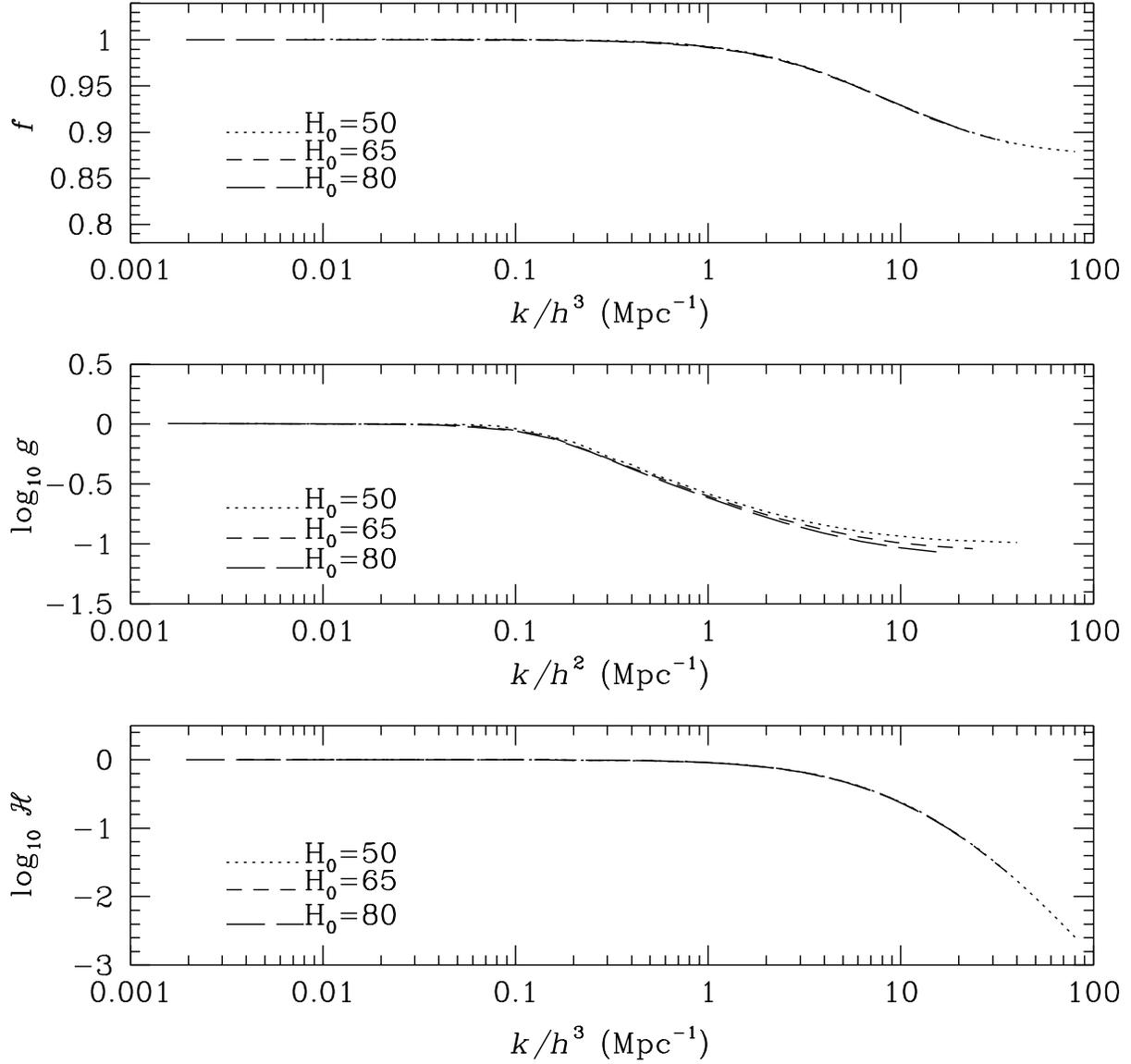}
\caption{Scaling of $k$ with the Hubble constant for functions $f$, $g$ 
and ${\cal H}$ in the $\onu=0.2$, $n=1$ CDM+HDM model at $a=1$.
As discussed in the text, $f$ and $\cal H$ scale perfectly 
with $k/h^3$, while $k/h^2$ is a good approximation for $g$ for
a large range of $k$.}
\label{fig:hscale}
\end{figure}
\begin{figure}
\epsfxsize=6.5truein 
\epsfbox{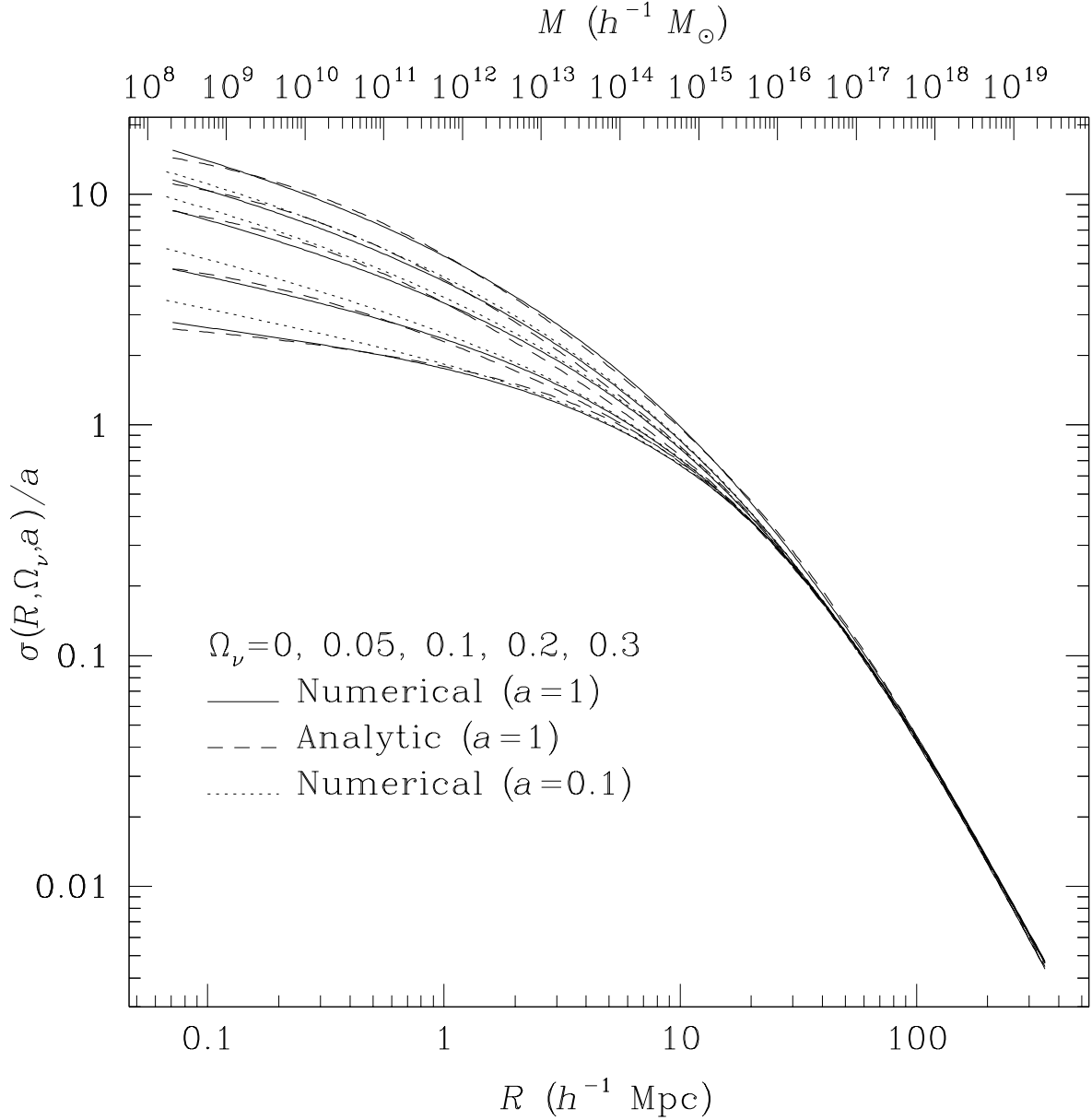}
\caption{Root-mean-square of the linear mass fluctuations $\sigma(R,\onu)$ 
in spheres of radius $R h^{-1}$ Mpc for $n=1$ and $H_0=50$ models with
$\onu=0.0, 0.05, 0.1, 0.2$, and 0.3 (top down).  The solid and dashed
curves are from numerical integration and the fitting formulas
eqs.~(\protect{\ref{sigfitc}}) and (\protect{\ref{sigfit}}),
respectively.  The dotted curves show the scaled
$a^{-1}\sigma(R,\onu\,,a)$ for $a=0.1$, which start to deviate from
$\sigma(R,\onu,a=1)$ only at small scale.  All models are normalized
to $\Qrms=18\,\mu K$ at large $R$. }
\label{fig:sig}
\end{figure}
\begin{figure}
\epsfxsize=6.5truein 
\epsfbox{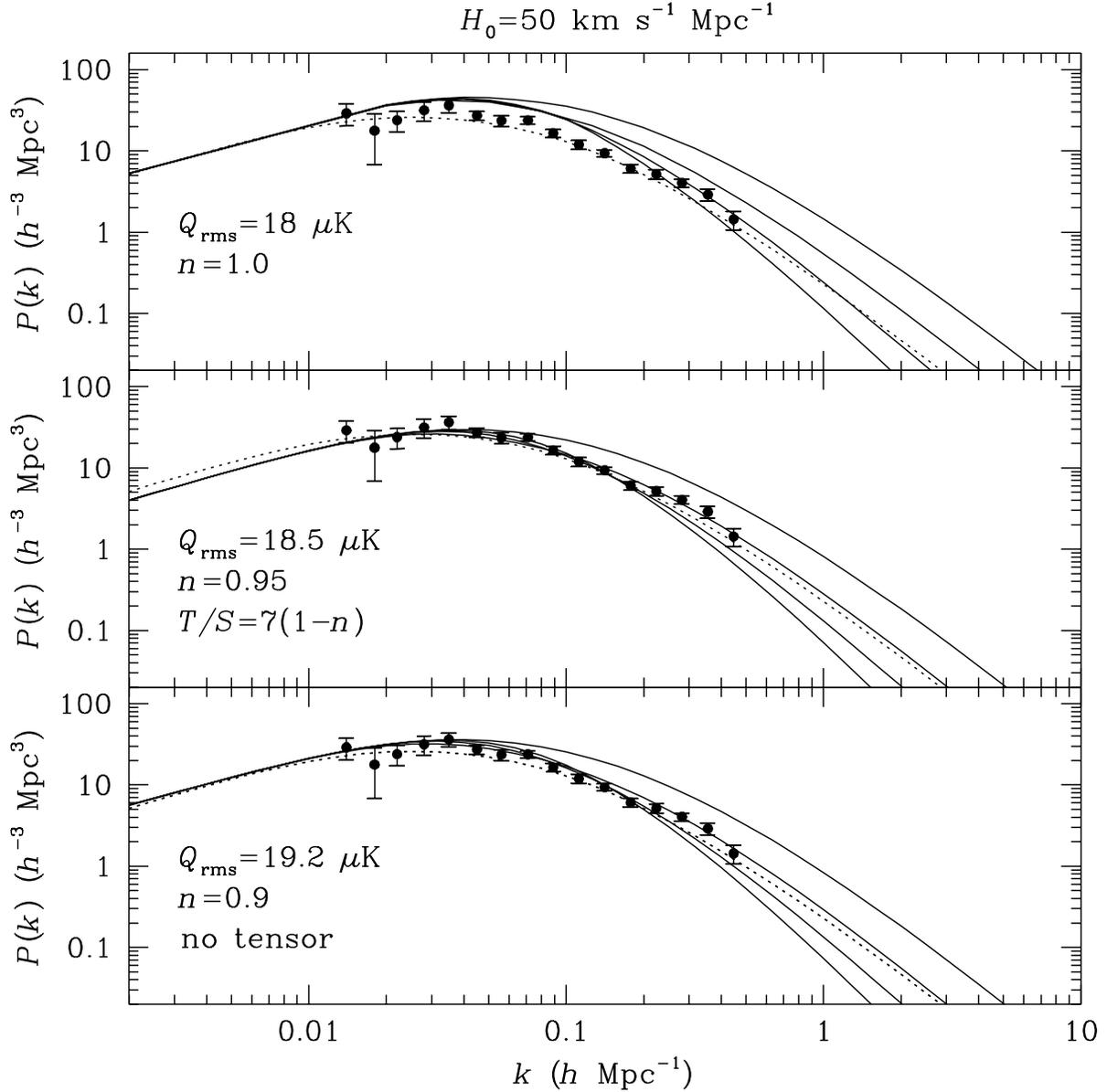}
\caption{Density-weighted linear power spectrum for four models with 
neutrino fractions $\onu=0.0, 0.1, 0.2$ and 0.3 (solid curves from top
down at large $k$), compared with the linear power spectrum
reconstructed by Peacock \& Dodds (1994) from galaxy and cluster
surveys (filled symbols).  The baryon fraction is $\Omega_b=0.05$ and
$H_0=50$ km s$^{-1}$ Mpc$^{-1}$.  The bottom two panels show the
excellent agreement of CDM+HDM models with the data when the spectral
index is reduced to $n=0.95$ with tensor fluctuations (middle) or to
$n=0.9$ without tensor fluctuations (bottom).  The 4-year COBE result
is used for the normalization, and the dependence of $\Qrms$ on $n$ is
taken into account (see text).  The dotted curves show an $n=1$
CDM-type spectrum with a shape parameter of $\Gamma=\Omega h=0.25$.}
\label{fig:powave}
\end{figure}
\begin{figure}
\epsfxsize=6.5truein 
\epsfbox{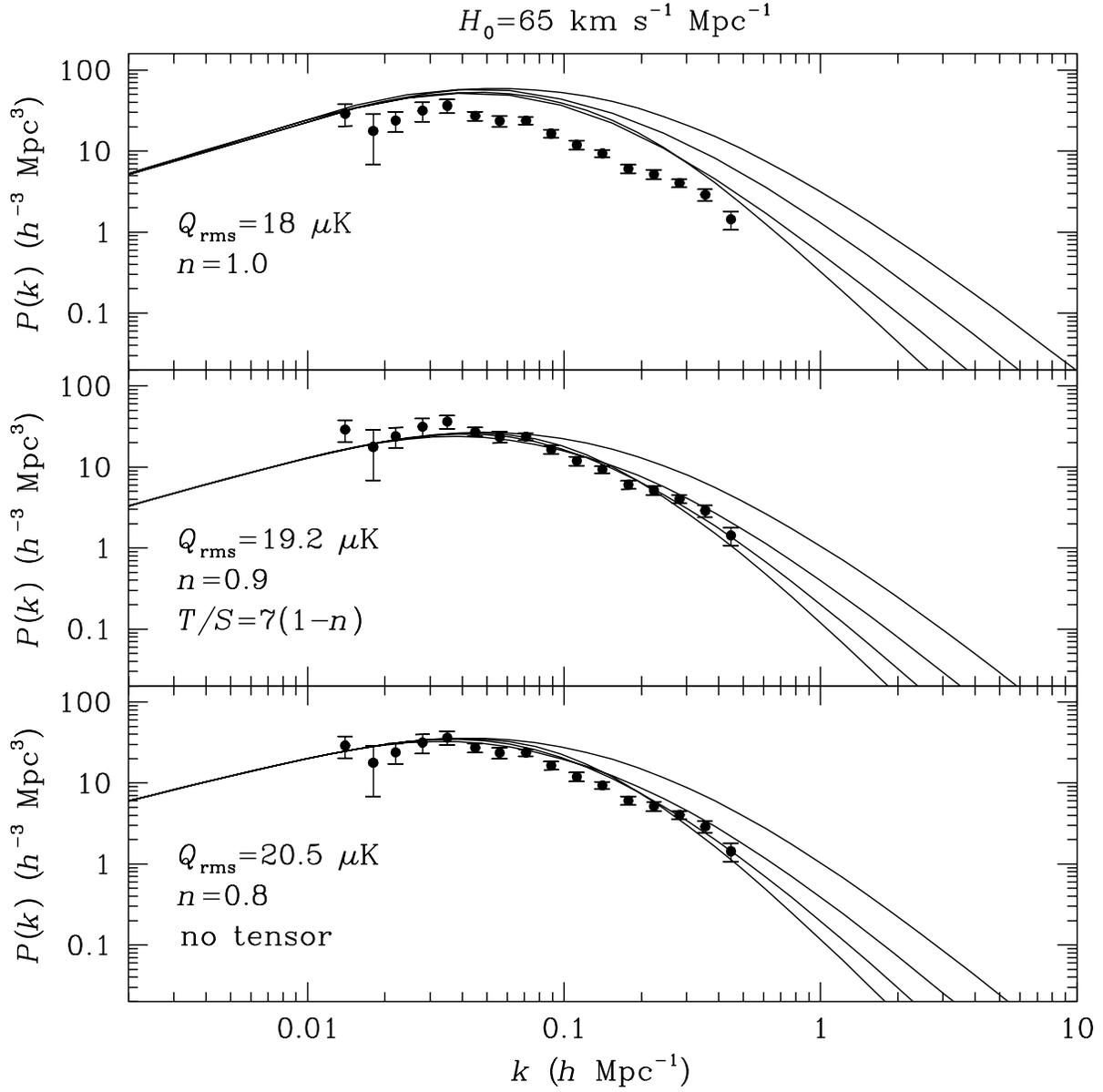}
\caption{Same as Figure~\protect{\ref{fig:powave}} but for $H_0=65$ km
s$^{-1}$ Mpc$^{-1}$ (with $\Omega_b=0.05$).  A larger tilt of
$n\approx 0.9$ (with tensor) or $n\approx 0.8$ (without tensor) is
needed to reduce the excess power in the $n=1$ model.}
\label{fig:powave65}
\end{figure}
\begin{figure}
\epsfxsize=6.5truein 
\epsfbox{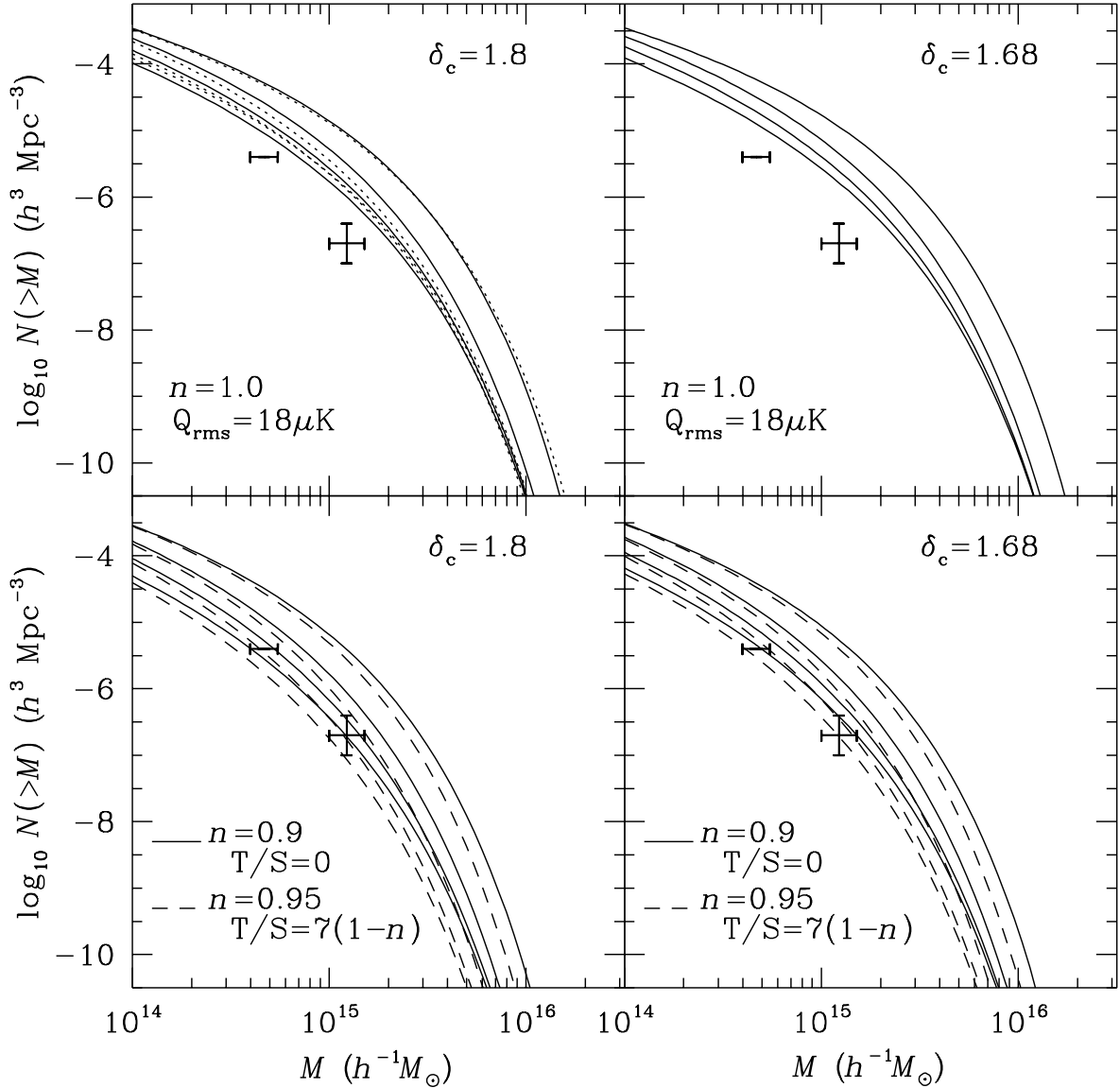}
\caption{Present-day comoving number density of cluster-mass objects
as a function of mass for the standard CDM and three CDM+HDM models.
Different panels illustrate the effect of varying the spectral index
$n$ and the parameter $\delta_c$ in the Press-Schechter approximation.
Top panels are for $n=1$ models with $\onu=0$, 0.1, 0.2, and 0.3 (from
top down), normalized to the 4-year COBE result $\Qrms=18\,\mu K$
(Gorski et al. 1996).  The solid and dotted curves in the upper left
panel compare $N(>M)$ computed from numerical vs. fitted $\sigma$.
Bottom panels show the same tilted models with
$\onu=0$, 0.1, 0.2, and 0.3 (from top down) as in
Figs.~\protect{\ref{fig:powave}} and \protect{\ref{fig:powave65}}.
The dependence of $\Qrms$ on $n$ is taken into account:
$\Qrms=19.2\,\mu K$ for $n=0.9$ and $\Qrms=18.5\,\mu K$ for $n=0.95$.
The data points indicate the number density of clusters (Henry \&
Arnaud 1991) with X-ray temperature exceeding 3.7 (upper left) and 7
keV (lower right), respectively.  The corresponding mass range is
taken from White et al. (1993a) and Liddle et al. (1996).}
\label{fig:cluster}
\end{figure}
\begin{figure}
\epsfxsize=6.5truein 
\epsfbox{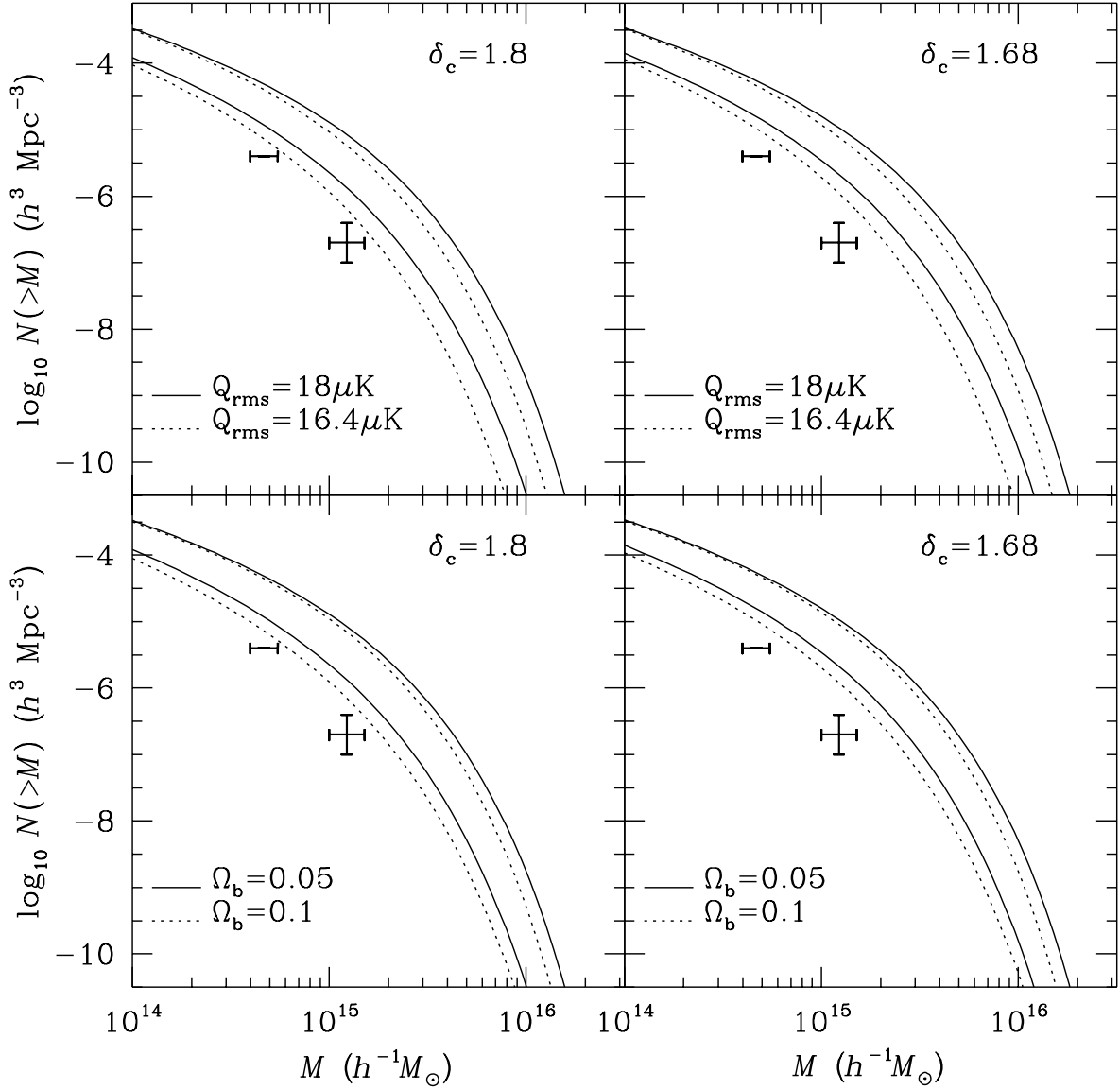}
\caption{Effects of changing $\Qrms$ (upper panels) and $\Omega_b$ 
(lower panels) on the number density of cluster-mass objects.  For
clarity, only $\onu=0$ and 0.3 models are shown.  All models have
$n=1$ and $h=0.5$.  The top panels assume $\Omega_b=0.05$; the bottom
panels assume $\Qrms=18\,\mu K$.  The data points are the same as in
Fig.~\protect{\ref{fig:cluster}}.}
\label{fig:cluster2}
\end{figure}

\end{document}